\documentclass[aps,prb,twocolumn]{revtex4}

\begin{document}
\title{
Reply to ``Comment on `First-principles calculation of the superconducting 
transition in MgB$_2$ within the anisotropic Eliashberg formalism'"
}

\author{
Hyoung~Joon~Choi,$^1$ David~Roundy,$^{1,2}$ Hong~Sun,$^1$ Marvin~L.~Cohen,$^{1,2}$ 
and Steven~G.~Louie$^{1,2}$
}

\address{
$^1$Department of Physics, University of California at Berkeley, Berkeley, California 94720\\
$^2$Materials Sciences Division, Lawrence Berkeley National Laboratory, Berkeley, California 94720
}

\date{\today}

\begin{abstract}
The recent preprint by Mazin {\em et al.} [cond-mat/0212417] contains many inappropriate evaluations 
and/or criticisms on our published work [Phys. Rev. B {\bf 66}, 020513 (2002) and 
Nature {\bf 418}, 758 (2002)]. The preprint [cond-mat/0212417v1] was submitted to Physical Review B 
as a comment on one of our papers [Phys. Rev. B {\bf 66}, 020513 (2002)]. 
In the reviewing process, Mazin {\em et al.} have withdrawn many of the statements
contained in cond-mat/0212417v1, however two claims remain in their revised manuscript 
[cond-mat/0212417v3]: (1) the calculated variations 
of the superconducting energy gap within the $\sigma$- or the $\pi$-bands are not observable 
in real samples due to scatterings, and (2) the Coulomb repulsion $\mu({\bf k},{\bf k'})$ 
is negligibly small between $\sigma$- and $\pi$-states and thus should be approximated 
by a diagonal $2\times 2$ matrix in the $\sigma$ and $\pi$ channels. Here, we point out 
that the former does not affect the validity of our theoretical work which is for the clean limit, 
and that the latter is not correct.
\end{abstract}

\maketitle

Our computational work\cite{1,2} for the superconducting properties of MgB$_2$ is an exact 
implementation of the fully anisotropic Eliashberg formalism which was established 
more than two decades ago. The momentum dependence and the dynamical behaviors of 
the electron-phonon interactions are fully considered in this formalism. 
By solving the anisotropic Eliashberg equations correctly without any assumption on 
the distribution of the superconducting energy gap on the Fermi surface, we conclusively 
obtained the theoretical values of the superconducting energy gap in MgB$_2$ 
for the first time\cite{2}.

Recently, Mazin {\em et al.} incorrectly stated in the original version of their 
preprint\cite{3} that our work\cite{1,2} represents a computational implementation of ideas 
proposed by Liu, Mazin, and Kotus\cite{4} and that our treatment of the fully anisotropic 
Eliashberg equations integrates out all the phononic degrees of freedom. 
These statements are deleted in the revised version of their preprint\cite{5}. 
Mazin {\em et al.} also incorrectly claimed in their preprint\cite{3} that the Coulomb 
pseudopotential between $\sigma$ and $\pi$ sheets might have been omitted erroneously 
in our actual computation. This incorrect claim is also deleted in their revised preprint\cite{5}.

The superconducting energy gap in MgB$_2$ is shown to have greatly different average 
values on the $\sigma$- and the $\pi$-sheets of the Fermi surface, and it also exhibits 
some variations within the $\sigma$- or the $\pi$-sheets\cite{2}. This is the result of {\em ab initio} 
calculations of Ref. 2 carried out within the fully {\bf k}-dependent (anisotropic) Eliashberg 
formalism in the clean limit, and no claim was made in our paper regarding whether the 
variations within the $\sigma$- or the $\pi$-sheets would be measurable in particular samples. 
We agree with the discussion\cite{3,5} that these variations inside 
the $\sigma$- or the $\pi$-sheets 
may be averaged out with sufficient strong scatterings in a sample, as also would eventually 
be the case for the difference between the $\sigma$ and $\pi$ gaps. Since this is an impurity 
effect, the theoretical result in Ref. 2 remains valid for samples in the clean limit. 
We hope that perhaps clever experiments in the future may observe these variations in 
appropriate samples.

Mazin {\em et al.} also claimed that the anisotropy of the Coulomb repulsion $\mu({\bf k},{\bf k'})$ 
on the Fermi surface is large and very important in MgB$_2$\cite{3,5}. They further argued that, when 
viewed as a $2\times 2$ matrix in the $\sigma$ and $\pi$ channels, $\mu$ should be taken 
as a diagonal matrix (i.e., no repulsion between $\sigma$ and $\pi$ states) as opposed to a uniform 
matrix in order to obtain physical results.  This claim was based on their simplified 
estimation of $\mu$ in MgB$_2$ by replacing the screened Coulomb interaction with 
a contact potential $\delta({\bf r}-{\bf r'})$. Their model calculation gives the result 
of $\mu_{\sigma\sigma} : \mu_{\pi\pi} : \mu_{\sigma\pi} : \mu_{\pi\sigma} =  3.1 : 2.6 : 1.4 : 1.0$.

A realistic calculation of $\mu$ requires knowing the screened Coulomb interaction 
$W({\bf r},{\bf r'})$ which typically has an extent of over a bond length. 
{\em Ab initio} calculations 
of $\mu$ can be achieved by employing the full dielectric matrix\cite{6}. This approach has 
been applied to MgB$_2$ recently and obtained a result\cite{7} of $\mu_{\sigma\sigma} : \mu_{\pi\pi}
: \mu_{\sigma\pi} : \mu_{\pi\sigma} = 1.75 : 2.04 : 1.61 : 1.00$ following the definition in Ref. 5.
These {\em ab initio} results show that $\mu$ is much less anisotropic 
than the model results in Ref. 5. Thus, the anisotropy of $\mu$ is not as 
important as claimed in Ref. 5, in particular in view of the fact that the value and 
anisotropy of $\lambda$ is an order of magnitude bigger in this system. In a simple modeling of 
$\mu$, it may in fact be more appropriate to take it as a uniform matrix than a diagonal one.

In addition, Figure 1 in Ref. 5 presents an inappropriate reduction of our full 
{\bf k}-dependent theory\cite{1,2} to a two-band model. It is likely that the claimed discrepancies 
originate from their incorrect splitting of the {\bf k}-dependent $\alpha^2 F(\omega)$ into 
$\alpha^2 F_{ij}(\omega)$'s for $i,j=\sigma,\pi$. (They used the same frequency dependence 
for all components -- $\sigma$-$\sigma$, $\pi$-$\pi$, and $\sigma$-$\pi$.) A two-band 
model properly reduced from our fully {\bf k}-dependent Eliashberg formalism in fact reproduces well 
the results in Refs. 1 and 2. These 2-band model results will be published elsewhere\cite{8}.

\end{document}